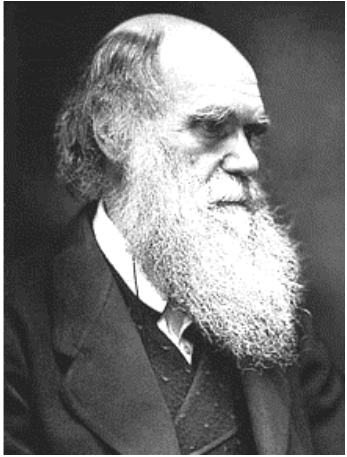
Charles Robert Darwin (1809-1882)

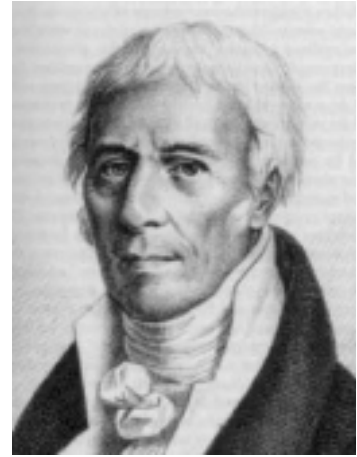
Jean-Baptiste Lamarck (1744-1829)

# Quantum mechanics as a sociology of matter[1]

Raoul Nakhmanson[2]

*Analogies between quantum mechanics and sociology lead to the hypothesis that quantum objects are complex products of **evolution**. Like biological objects they are able to receive, to work on, and to spread semantic information. In general meaning we can name it "Consciousness". The important ability of consciousness is ability to predict future.*

Key words: Evolution, consciousness, information, quantum mechanics, EPR, decoherence.

The term "sociology" appears first in the middle of the 19th century. Dictionaries and encyclopedias define sociology as a science for society and self-contained social institutions. At the beginning it was related to humans only, but in the 20th century sociologic researches were made with animals and insects too.

Between quantum mechanics and sociology there are analogies:

1) Like sociology, quantum mechanics describes societies ("ensembles") as a whole;

2) In respect to individuals (people or particles) there are only probabilities;

3) Members belonging to some society or quantum ensemble being under investigation are similar to each other but can be very different from members of other societies or ensembles (species in biology as well as in "zoology" of particles);

4) Members belonging to some society or ensemble are regarded not only as similar but as identical in the sense that exchange of any two members does not alter the ensemble. In quantum mechanics it guided to symmetrical and antisymmetrical wave functions belonging to bosons and fermions having collective and anti-collective behavior, respectively. In sociology one also find such two type of behavior.

5) The evolution of societies is subjected to some intention. In quantum mechanics it is seen in the least action principle;

6) Members of societies are "atoms" and "individuals" literally, that is, from one person or one electron it is impossible to make two smaller persons or two smaller electrons.

---

[1] Theses for IC QT RF 2, Växiö 2003.
[2] E-mail: nakhmanson@t-online.de

The last item prompts us that the members of biological and quantum societies are complex constructions penetrated with inside connections. Their decomposition destroys their functioning.

Complex biological objects develop themselves during evolution. To begin the analogous evolution process with matter some "dark" matter in non-equilibrium state is needed. Such a state can be the result of a fluctuation. Modern physics operates with sizes up to $10^{-35}$ meter (Planck length). If evolution of matter began 14 billions years ago, the elementary particles having sizes $10^{-15} - 10^{-20}$ m have a long evolution time behind and perhaps do not yield up to biological objects in complexity and functioning.

From the example of biological objects we know that the essence of life is information. The ability to receive, to work on, and to spread information is important for individuals and societies, and is selected by evolution. We speak about individual and social consciousness developing in parallel with material structures.

Evolution theory and analogy between quantum mechanics and sociology lead to the idea that material objects of quantum mechanics have an individual and a social consciousness too. Such a hypotheses explains the essence of wave function and its collapse as well as "mysterious" experiments known as "two-slits", "delayed-choice", "EPR", "Aharonov-Bohm", and "interaction-free measurement". The important ability of consciousness is ability to predict future. Here also are roots of entanglement of parted particles. The depth of prediction is limited by interaction with environment, thereafter comes decoherence.

Transition from quantum to classical physics is thought as a transition from individuals to crowds. Crowds are "dividual": One crowd can be split in two smaller crowds. Crowds' behavior is more deterministic than the behavior of individuals.

In such a context the wave function is a purely mental construction in an abstract configuration space which, in its turn, is in a consciousness of microparticle. If you allow me a pun, the so-called "waves of matter" are non-material. Einstein justly called them "Gespensterfelder". Nevertheless they control the behavior of material objects. Of course physicists can have in their mind not only their own wave function (it is strategy of their behavior) but also some idea, correct or not, about wave function of the particle.

The wave-particle duality is a mind-body one. In the real 3D-space there exists only the particle, the wave exists in its consciousness. If there are many particles, their distribution in accordance with the wave function represents a real wave in real space. Many worlds, Schrödinger cat, Great Smoky Dragon, etc. exist only as virtual mental constructions.

Sociology and psychology interview persons being under investigation. The hypotheses of evoluting matter allows us to do it with quantum objects too. The consequent experiments include action on particles and atoms with semantic information.

Some details can be found in:
1. http://arXiv.org/pdf/physics/0004047
2. http://arXiv.org/pdf/physics/0111109
3. http://www.agharta.net/Superstrings.html  (in Russian)